\documentclass[sigconf, nonacm]{acmart}

\newcommand\vldbdoi{XX.XX/XXX.XX}

\newcommand\vldbavailabilityurl{URL_TO_YOUR_ARTIFACTS}

\usepackage{xcolor}
\usepackage{subcaption}
\usepackage{multirow}
\usepackage{listings}
\usepackage{algpseudocode}
\usepackage{enumitem}

\usepackage[linesnumbered,ruled,vlined]{algorithm2e} 

\begin{document}
\title{Reverse Migration of Cloud Applications to On-premises}

\author{Alekh Jindal,
Jyoti Pandey,
Christina Pavlopoulou,
Ronith PR,
Sharath Prakash,
Shi Qiao,
Shivani Tripathi,
Wangda Zhang}
\email{research@tursio.ai}
\affiliation{%
  \vspace{0.2cm}
  \institution{Tursio}
  \city{Bellevue}
  \country{USA}
  \vspace{0.2cm}
}

\begin{abstract}
Cloud computing has become ubiquitous for modern applications due to its agility and scalability. However, regulated industries still prefer on-premises deployments for security and compliance reasons. This creates a paradox for vendors who need to develop in the cloud but deploy on-premises, leading to long release cycles and complex maintenance.
In this paper, we describe Diel, the Tursio On-premises migrator framework to automate reverse migration of cloud applications to on-premises environments. Diel applies a combination of \textit{simulate}, \textit{replicate}, and \textit{delegate} strategies to transform cloud services into on-premises counterparts. We describe the design and implementation of Diel, along with lessons learned from using it in practice. With Diel, we have been able to keep Tursio AI's cloud and on-premises versions in sync, releasing new stable versions every three weeks.
\end{abstract}

\maketitle



\section{Introduction}

Cloud computing has become the de-facto platform for modern applications, offering unmatched scalability, availability, and cost-effectiveness~\cite{berkeley-cloud-computing-survey,cloud_computing_armbrust}. Yet, many enterprises---particularly those in regulated industries such as finance, healthcare, and government---still prefer on-premises deployments to meet stringent data security, compliance, and latency requirements. A recent Rackspace survey reveals that two-thirds of respondents have \textit{considered repatriating a portion of their workloads from public clouds back to private clouds or on-premises infrastructure}~\cite{rackspace-2025-state-cloud-report}. At the same time, cloud offers significant advantages for software development, including faster iteration cycles, streamlined CI/CD pipelines, and access to managed services.

This creates a dilemma for vendors serving regulated industries: they must either forgo the benefits of cloud-native development or maintain separate codebases for cloud and on-premises deployments, painfully porting changes between them. Both options are costly---the former sacrifices development velocity, while the latter leads to long release cycles and complex maintenance. The result is that many vendors are disincentivized from building on-premises products altogether, leaving regulated customers underserved.

In this paper, we describe \textit{Diel}, the Tursio On-premises framework for automating the reverse migration of cloud applications to on-premises environments. Diel identifies cloud-specific components and generates equivalent on-premises configurations, supporting the entire application stack---compute, storage, databases, workflow, logging, authentication, and even AI services. The key idea is to apply a combination of three strategies to transform cloud services into on-premises counterparts:
(1)~\textit{simulate}---creating local versions of cloud services that mimic their behavior;
(2)~\textit{replicate}---setting up on-premises instances using open-source or commercial alternatives; and
(3)~\textit{delegate}---routing certain functionalities back to the customer's cloud subscription when on-premises alternatives are not feasible.

\begin{figure}[!t]
  \includegraphics[width=0.475\textwidth]{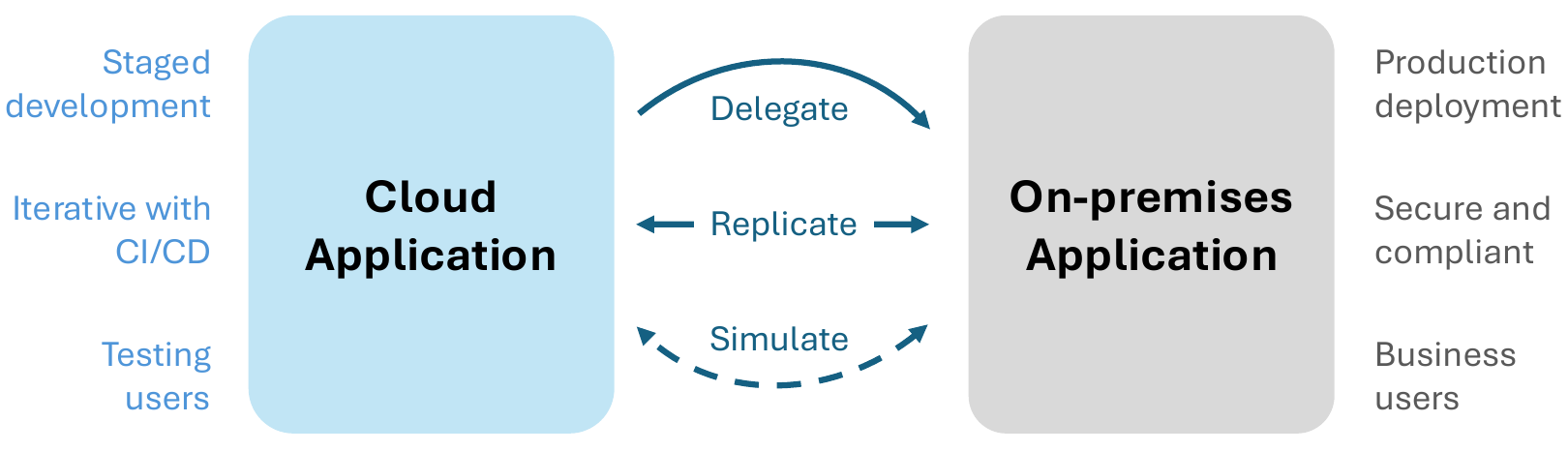}
  \caption{Diel bridges cloud-native development (left) with secure on-premises deployment (right) using simulate, replicate, and delegate strategies.}
  \Description{On-prem migrator illustration}
  \label{fig:onprem-migrator-Illustration}
\end{figure}

Figure~\ref{fig:onprem-migrator-Illustration} illustrates the concept of Diel. The left side shows the cloud application, under active development with rapid CI/CD iterations. The right side shows the on-premises deployment, running in production with full security and compliance. Diel maintains the bridge between these two worlds, enabling vendors to develop cloud-natively while deploying on-premises seamlessly.

\begin{figure*}[!t]
  \includegraphics[width=0.95\textwidth]{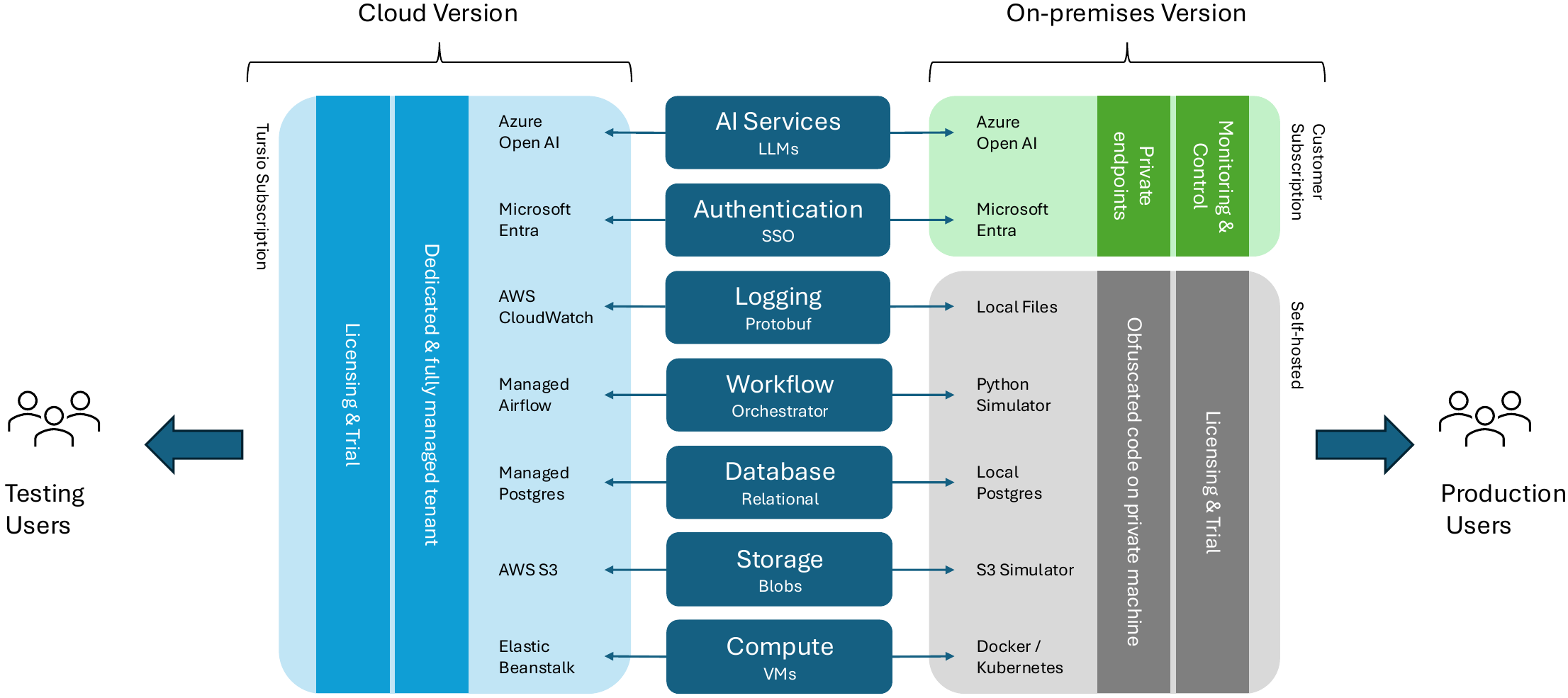}
  \caption{Diel reverse migration for the Tursio application stack: cloud services (blue, left) are simulated, replicated, or delegated to produce the on-premises deployment (gray/green, right).}
  \Description{On-prem migrator architecture}
  \label{fig:onprem-migrator}
\end{figure*}

With Diel, we have been able to keep Tursio AI's cloud and on-premises versions in sync, releasing a new stable version every three weeks. This cadence provides sufficient time to deliver new features and improvements to on-premises customers while maintaining quality. The customer feedback has been positive:

\begin{quote}
  ``Fastest AI implementation that I have heard of.'' -- VP at a Community Financial Institution.
\end{quote}

The rest of the paper is organized as follows. Section~\ref{sec:background} provides background on cloud migration and the challenges of on-premises deployments. Section~\ref{sec:reverse-migration} describes the design and implementation of Diel, including the simulate, replicate, and delegate strategies. Section~\ref{sec:lessons} discusses lessons learned from using Diel in practice. Section~\ref{sec:related-work} covers related work. Finally, Section~\ref{sec:conclusion} concludes the paper and outlines future work.

\section{Background}
\label{sec:background}

Cloud migration has been a central pillar of enterprise modernization over the last decade and a half, driven by cost savings, scalability, agility, and access to advanced services such as AI/ML and big data analytics. While privacy and security concerns were significant early on, cloud technology has gradually matured to address them. Yet, organizations in regulated industries still retain a significant portion of their data on-premises---and with data now being fed into AI systems, they find themselves at a crossroads once again. To minimize risk and accelerate AI deployment, keeping AI on-premises is often the most practical path forward.

At the same time, cloud has become the dominant platform for software development itself, with CI/CD pipelines, DevOps practices, and cloud-native architectures now the norm. This creates a paradox for vendors serving regulated industries: they need to develop in the cloud but deploy on-premises. Traditionally, this has led to long release cycles with tedious packaging and testing, a counterproductive approach in the fast-paced AI era, where users expect rapid innovation and frequent updates that are hard to achieve with traditional on-premises deployments.
The goal of Diel is therefore to maintain a near-live mirror from a cloud application to its on-premises counterpart.

\section{Reverse Migration}
\label{sec:reverse-migration}

Figure~\ref{fig:onprem-migrator} shows the Diel reverse migration for the Tursio application stack. The left side illustrates the cloud version (in blue), which hosts all services within the Tursio subscription. Tursio creates a dedicated tenant for each workload to ensure data isolation and access control.
The right side shows the on-premises version (in gray and green), managed by the customer. The gray portion is self-hosted, while the green portion is delegated back to the customer's cloud subscription.

To reverse-migrate the cloud application to on-premises, Diel:
(1)~\textit{simulates} compute, storage, and workflow services locally;
(2)~\textit{replicates} database and logging services using open-source alternatives; and
(3)~\textit{delegates} authentication and AI services back to the customer's cloud subscription, as these services are too complex or too costly to replicate or simulate locally.
Together, these services form the typical application stack for modern cloud applications. Tursio encrypts all code shipped on-premises for IP protection and encrypts data at rest and in transit, in both cloud and on-premises versions, for end-to-end security.

Below, we describe each of the three strategies in more detail.

\subsection{Simulate}

We simulate three core cloud services locally on-premises:

\subsubsection*{1. Compute.} The cloud version of the application manages compute using Elastic Beanstalk on AWS~\cite{aws-elastic-beanstalk}; other cloud providers offer similar services. For on-premises, we simulate compute locally using Docker and Kubernetes to ensure compatibility with the cloud environment. Our Docker image is built to match the cloud environment as closely as possible, including the same operating system, libraries, and dependencies. Optionally, we support Kubernetes to manage scaling, load balancing, and service discovery~\cite{docker-kubernetes}.

\subsubsection*{2. Storage.} The cloud version uses AWS S3~\cite{aws-s3} as the underlying storage for model and data files. For on-premises, we simulate S3 locally by implementing a local file system stub for the relevant Boto3 SDK methods, the API we use to access S3 in the cloud. This is similar to open-source S3 simulators such as MinIO~\cite{minio} and LocalStack~\cite{localstack}, but we opted for a lightweight custom implementation that fits our specific needs and can be extended as needed.

\subsubsection*{3. Workflow.} We use Airflow as our orchestration engine and employ AWS MWAA (Managed Workflows for Apache Airflow)~\cite{aws-mwaa} in the cloud version. For on-premises, we simulate Airflow locally using a combination of cron jobs to handle submission, monitoring, failure detection, and retry logic. The status of all background jobs is surfaced in the Tursio portal and remains consistent across both cloud and on-premises versions. Alternatively, Airflow can be deployed locally using Docker or Kubernetes, similar to the compute service; however, we prefer the lightweight cron job approach for simplicity and ease of maintenance.

\vspace{0.2cm}
By simulating these services locally, we maintain a consistent application stack that can be developed freely, without worrying about porting environments or managing them separately.

\subsection{Replicate}

We replicate two cloud services on-premises using open-source alternatives:

\subsubsection*{4. Database.} Tursio uses AWS RDS (Relational Database Service)~\cite{aws-rds} with PostgreSQL as the underlying database engine in the cloud version. For on-premises, we replicate the database using PostgreSQL directly, which is open-source and can be installed and managed locally. We ensure that the database schema, indexes, and data types remain consistent between the cloud and on-premises versions to minimize compatibility issues.

\subsubsection*{5. Logging.} The cloud version uses AWS CloudWatch~\cite{aws-cloudwatch} for logging and monitoring. For on-premises, we direct logs to stdout and collect them in the Docker container. Logs are rotated periodically, similar to CloudWatch, and compressed when the volume of log files grows large.
The compressed files retain only key metrics in protobuf format, without additional traces, to minimize the local storage overhead.

\vspace{0.2cm}
By replicating these two services locally, we support them on-premises seamlessly and with minimal additional effort.

\subsection{Delegate}

Finally, we delegate the remaining two cloud services back to the customer's cloud subscription:

\subsubsection*{6. Authentication.} Tursio uses Azure AD (Active Directory), now renamed to Microsoft Entra ID~\cite{microsoft-entra-id}, for user authentication and management in the cloud version. For on-premises, we delegate authentication back to the customer's Azure AD tenant, allowing users to log in with their existing credentials. This ensures that user management remains centralized and consistent across both cloud and on-premises versions.

\subsubsection*{7. AI Services.} Tursio leverages large language models (LLMs), specifically the OpenAI~\cite{openai} models, in the cloud. For on-premises, we delegate AI service calls back to the customer's Azure OpenAI~\cite{azure-ai-foundry-openai-models} subscription, allowing the application to access the same AI capabilities while maintaining data security, compliance, and content filtering via the customer's own Azure subscription. If a customer hosts private models (e.g., Llama, Qwen, or others), Tursio can also be configured to use those internal model endpoints instead.

\vspace{0.2cm}
Authentication and LLMs are key enterprise components; delegating them back to the customer's cloud ensures that existing security and compliance policies remain intact.

\section{Lessons Learned}
\label{sec:lessons}

In this section, we share lessons learned from the Diel reverse migration at Tursio.

\subsubsection*{Testing is still a critical step.} While Diel automates much of the migration process, thorough testing remains essential to ensure that the on-premises version functions correctly. The bar for on-premises software quality is very high, and once installed, it is difficult to recall or patch quickly. Hence, we dedicate a one-week testing phase, including integration and user acceptance tests, before releasing each new on-premises version.

\subsubsection*{IT teams can be the bottleneck.} On-premises deployments often require coordination with the customer's IT team, which can introduce delays and complications. To partly mitigate this, we provide detailed documentation and support to help IT teams understand the deployment process and address potential issues upfront. Yet, the dependency on an expert human in the loop remains an unavoidable challenge.

\subsubsection*{Debugging is a challenge.} Debugging on-premises deployments can be more challenging than cloud deployments due to limited access to logs and monitoring tools. To address this, we have implemented robust logging and monitoring mechanisms that allow us to collect and analyze logs from on-premises deployments effectively. We also provide customers with tools to export and share logs securely. However, debugging remains a time-consuming process, as it requires someone on the customer side to share logs before Tursio can investigate further.

\subsubsection*{Upgrading seamlessly.} Diel accelerates new releases for on-premises customers, but the upgrade process itself still needs to be seamless. We have developed a robust upgrade mechanism that ensures smooth transitions between versions, minimizing downtime and disruption. This includes backup and rollback strategies in case of issues during the upgrade. However, the upgrade process is still not as quick as cloud deployments, and upgrades must be planned carefully to minimize impact on customers.

\subsubsection*{Conforming to custom security and compliance.} Organizations often have custom security and compliance requirements that must be addressed in on-premises deployments, e.g., vulnerability scanning, AI governance, data residency, and more. This requires us to provide extensibility stubs where customers can plug in their own endpoints and policies. Unfortunately, such customizations require significant back-and-forth and are time-consuming. We handle them on a case-by-case basis.

\subsubsection*{Product training is important.} On-premises deployment means that the Tursio team cannot have direct access to the customer environment. Thus, we train the customer team using our cloud environment (even providing access to a test Tursio cloud instance) so that they can operate and manage the on-premises application effectively. Obviously, this requires time commitment from the customer team, which can be a challenge for some organizations.

\vspace{0.2cm}
In summary, Diel has been instrumental in enabling Tursio to deliver cloud-native applications on-premises efficiently. While challenges remain, the benefits of automated reverse migration far outweigh the complexities involved.

\section{Related Work}
\label{sec:related-work}

\subsubsection*{Cloud migration.}
A large body of work addresses migrating applications \textit{to} the cloud. Jamshidi et al.~\cite{jamshidi2013cloud} provide a systematic review of cloud migration research, cataloging strategies such as rehosting, replatforming, and refactoring. Balalaie et al.~\cite{balalaie2016microservices} describe patterns for incrementally migrating monolithic applications to cloud-native microservices architectures. Kratzke and Quint~\cite{kratzke2017cloud-native} present a systematic mapping study of cloud-native application research after a decade of cloud computing. These works focus on the forward direction---moving to the cloud---whereas Diel addresses the reverse: migrating cloud applications back to on-premises.

\subsubsection*{Cloud repatriation.}
The trend of moving workloads back from the cloud has gained attention in recent years. Wang and Casado~\cite{wang2021cost-of-cloud} argue that at scale, cloud costs can significantly erode margins, motivating repatriation. Dropbox's migration of over 600\,PB of data from AWS to its custom on-premises storage system~\cite{dropbox-magic-pocket} is one of the most prominent real-world examples. Industry surveys confirm the trend: the Rackspace report~\cite{rackspace-2025-state-cloud-report} finds that two-thirds of respondents have considered repatriating workloads. However, these efforts are largely ad-hoc and infrastructure-focused. Diel differs by providing a systematic, application-level reverse migration framework with reusable strategies.

\subsubsection*{Cloud portability and abstraction.}
Several efforts aim to reduce cloud vendor lock-in~\cite{opara2016vendor-lock-in} through portability standards and abstraction layers. TOSCA~\cite{binz2014tosca} defines a portable topology specification for deploying applications across heterogeneous cloud environments. Container technologies~\cite{pahl2019cloud-containers} such as Docker and Kubernetes enable compute portability, while Infrastructure as Code tools~\cite{morris2020iac} like Terraform and Pulumi allow declarative provisioning across providers. Cloud emulators such as LocalStack~\cite{localstack} and MinIO~\cite{minio} simulate individual AWS services for local development. These tools address specific layers of the stack---infrastructure provisioning, compute portability, or individual service emulation---but they don't provide a reverse migration of a full application stack. Diel fills this gap by combining simulate, replicate, and delegate strategies to transform an entire cloud application into an on-premises deployment.

\section{Conclusion and Future Work}
\label{sec:conclusion}

Cloud has become the default platform for modern applications, yet regulated industries still require on-premises deployments for data security, compliance, and latency reasons. Developing and maintaining separate codebases for cloud and on-premises environments, however, is time-consuming and complex.
In this paper, we presented Diel, the Tursio On-premises Migrator, which automates reverse migration of cloud applications to on-premises environments using a combination of simulate, replicate, and delegate strategies. With Diel, we have been able to develop Tursio as a cloud application while shipping it on-premises seamlessly, keeping both versions in sync with regular releases.

Looking ahead, we plan to extend the Diel approach and make Tursio available in various cloud marketplaces, including Microsoft Marketplace (already in preview), AWS Marketplace, and Google Cloud Marketplace. This will further simplify the deployment process for on-premises customers, reduce the dependency on their IT teams, and expand our reach to a broader audience. Additionally, we aim to enhance the Diel toolset with more advanced features, such as automated compliance checks and security audits, to further streamline the on-premises deployment process. Our goal is to make on-premises deployments as seamless and efficient as cloud deployments, so that regulated industries can benefit from both security and speed.



\begin{acks}
We thank our colleagues at Tursio for their contributions to building and refining Diel, and our on-premises customers for their valuable feedback throughout the deployment process.
\end{acks}

\balance
\bibliographystyle{ACM-Reference-Format}
\bibliography{references}

\end{document}